\documentstyle[aps,epsf]{revtex}
\begin{document}
\draft
\title{Density mismatch in thin diblock copolymer films}
\author{S. Martins$^1$, W.A.M. Morgado$^1$, M.S.O. Massunaga$^2$ and M. Bahiana$^1$}
\address{$^1$Instituto de F\'\i sica, UFRJ, Caixa Postal 68528, Rio de Janeiro, RJ, Brazil, 21945-970\\
$^2$Laborat\'orio de Ci\^encias F\'\i sicas, Universidade
Estadual
do Norte Fluminense, \\Av. Alberto Lamego 2000,
Campos dos Goytacazes, RJ, Brazil, 28015-620}

\date{\today}
\maketitle
\begin{abstract}
Thin films of diblock copolymer subject to gravitational field
are simulated by means of a cell dynamical system model.
The difference in density of the two sides of the molecule
and the presence of the field causes the formation of
lamellar patterns with orientation parallel to the confining
walls even when they are neutral. The concentration profile of
those films is analyzed in the weak segregation regime and a
functional form for the profile is proposed.

\end{abstract}
\section{Introduction}
Recently, there has been a great deal of interest in problems involving
periodic patterns in the tens of nanometers scale, for example,  
light conduction by photonic
crystals\cite{joanopoulos}, Josephson junctions arrays 
formed by granular superconducting 
materials\cite{relevantes}, and lithographic masks for special design chips 
\cite{park1} among others.

In special, nanolithography has brought considerable attention
to thin films  of microphase separated diblock copolymers (DBCP) 
as they naturally self organize in periodic structures on that length scale \cite{park1,PhTo,zhong}.

The basic
technique for the fabrication of those templates is the creation of
a well-ordered DBCP film between two flat surfaces and then the transfer of
the
microdomains to a substrate where one of the components is removed 
leaving
behind an ordered array of stripes or dots in either low or high
relief. The desirable ordering, in this case, is one that creates a
pattern
on the substrate; for example, for even DBCP molecules one must have
lamellae perpendicular to the hard boundaries. 

For that matter it is important to understand the pattern
formation in confined films of DBCP since it involves problems 
not present in bulk systems. This issue 
has been addressed both theoretically
\cite{turner1,kiku,brown-chakra2,pibal1} and experimentally
\cite{menelle,lamb,kellogg1} and the basic conclusions are 
that, when the confining walls
are neutral, the equilibrium pattern corresponds to lamellae
perpendicular
to the walls and, when 
the substrate prefers one kind of monomer, the pattern
may consist of lamellae parallel or perpendicular to the walls,
depending
on the relation between the film thickness and the bulk lamellar width.
The
later effect appears because the finiteness of the system brings about
frustration and one has to take into account the amount of compression
or
stretching of the molecules in order to accommodate a certain number of
lamellae between the two rigid walls.

One important issue that has not been emphasized in the above studies 
is the possibility of density mismatch
between the two parts of the molecules. At the bottom
wall, as the denser part of
the molecule sinks, lamellae parallel to the walls will form, even if
the
walls are neutral \cite{canela,bamorg1}. Being a bulk effect, the
interaction with the gravitational field
is capable of dramatically changing the microphase separation, even for 
infinite systems, as the lamellae tend to be aligned with the field
far from the boundaries
\cite{canela}. In finite systems, the lamellae perpendicular to the
field
present more diffuse interfaces and the gravitational field may
completely
destroy the microphase separation \cite{bamorg1}.
In the present work we consider this problem on two-dimensional films
of even DBCP
molecules as we analyze the effects of the degree of polymerization and film
thickness on frustration by means of a cell dynamical
system(CDS) model. We simulate films both in the weak and strong
segregation regimes ((WSR)and (SSR)). For the WSR we empirically find
the one-dimensional concentration profile and study the distortion of each 
layer within the lamellae.

In Sec. II, we define the model and outline the numerical scheme.
Results for 
neutral and interacting walls in the presence and absence of
the gravitational field are discussed in Sec. III. The effects of
frustration are analyzed as they affect not only the size and number of lamellae,
but also their internal structure.
In section V, the main conclusions are summarized.

\section{Computational Model}
Block copolymers are linear-chain molecules 
consisting of two subchains $A$ and $B$ grafted covalently to 
each other. Below some critical temperature $T_c$ these two blocks
 tend to separate, but due to the covalent bond, they can segregate at best 
locally to form periodic structures \cite{PhTo}. Here we consider only
the case of even molecules corresponding to lamellar equilibrium
patterns.
CDS models have been successfully used in several problems of phase
separation dynamics
due to their computational efficiency and versatility
\cite{oopu1,enokawa,chakragun1,baoo,mogo1,shinooo2}, so we prefer that
method for the simulations.
 As usual, in this kind of 
description 
 we assign a scalar variable $\psi(n,t)$ to each lattice site 
corresponding to the coarse-grained order parameter in the $n$-th cell
at time
 $t$ (time here is defined as the number of iterations). This order
parameter 
represents the difference $\psi_A-\psi_B$, where $\psi_A(\psi_B)$ is the
local
 number density of $A(B)$. The ingredients for the time evolution of
$\psi$ 
are: local dynamics dictated by a function with two symmetric hyperbolic 
attractive fixed points, diffusive coupling with neighbors,
stabilization of 
the homogeneous solution and conservation of $\psi$. For the present
problem, we also
add the interaction with the gravitational field and with the confining
walls. The conservation, when an
 external field is present, must be imposed by considering the Kawasaki 
exchange dynamics explicitly. The detailed explanation of this model is
found in \cite{kitaoo} for spinodal decomposition. With this, we come to
final 
equation for a melt of even DBCP molecules:
\begin{equation}
\psi(n,t+1)=(1-\epsilon)\psi(n,t)+
\left\langle\left\langle C\left(n,j;sgn [I(n,t)-I(j,t)]\right)
 [I(n,t)-I(j,t)]\right\rangle\right\rangle,  
\label{eq-model}
\end{equation}
where 
\begin{equation}
I(n,t) \equiv {\cal A} \tanh\left(\psi(n,t)\right)-\psi(n,t)+
D\left[\langle\langle \psi(n,t)\rangle\rangle-\psi(n,t)\right]+hn_z+
V_s(n)
\end{equation}
is essentially the chemical potential.
$\langle\langle\star\rangle\rangle$
 is the isotropic space average of $\star$, ${\cal A}$ is 
a measure of the quench depth, and
$D$ is the diffusion coefficient. The parameter $\epsilon>0$ 
appears in this model to 
stabilize the solution $\psi=0$ in the bulk, for $\epsilon=0$ we have a
model
for 
spinodal decomposition, in which the domains can grow without bound.
Scaling
 arguments have proved that $\epsilon\sim N^{-2}$, where $N$ is the
polymerization 
index \cite{ooba1}. $h$ is the gravitational field, which we assume is
in
the $z$
 direction, and $n_z$ is the $z$ component of $n$. For molecules with
matched densities we
just take $h=0$. A possible interaction with the walls appears via the
surface term $V_s(n)$. $C$ is the collision
coefficient  given by: 
$C(i,j;\alpha)=[\psi_c+\alpha\psi(j)][\psi_c-\alpha\psi(i)]/\psi_c^2$,
where 
$\pm\psi_c$ are the fixed points of ${\cal A}\tanh\psi -\psi$ for ${\cal A}>1$.
For all the simulations  we used the values ${\cal A}=1.2$ and $D=0.5$, 
and uniformly distributed random initial conditions. The gravitational
field,
when 
present, is parallel to the smaller dimension. The
direction
 normal to the field will be called the $x$ direction.
We consider systems with periodic boundary conditions in the $x$ direction and
hard
walls in the $z$ direction, separated by a distance $L_z$. At the hard
walls we impose no flux boundary 
conditions in the form:$[I(z+1)-I(z)]_{boundaries}= 0$. 
\section{Results}
%
%
In order to understand the effect of confinement on the lamellae width,
we
must first determine its bulk value. For that matter, 
we ran simulations on $512\times 512$ lattices with periodic boundary
conditions for different values of $\epsilon$, and $h=0$. The resulting
isotropically
striped pattern was then Fourier
transformed, and the bulk lamellar width $W_b$ was measured in a
standard
way. Defining one
lamella as  $ABBA$ we have:
\begin{equation}
W_b = \frac{2\pi}{\langle k\rangle_{eq}},
\end{equation}
where
\begin{equation}
\langle k\rangle_{eq} = \frac{\int S(k,\infty)\,k\, {\rm d}k}{\int
S(k,\infty)}.
\end{equation}
$S(k,\infty)$ is the circular average of the structure function 
$S({\bf k },t)=|\psi({\bf k},t)|^2$, calculated at large times, that is,
when the value of 
$\langle k\rangle$ approaches a constant value. 

\begin{figure}
\setlength{\unitlength}{1mm}
\begin{picture}(90,90)(0,0)
\put(50,5){\epsfxsize=8cm\epsfbox{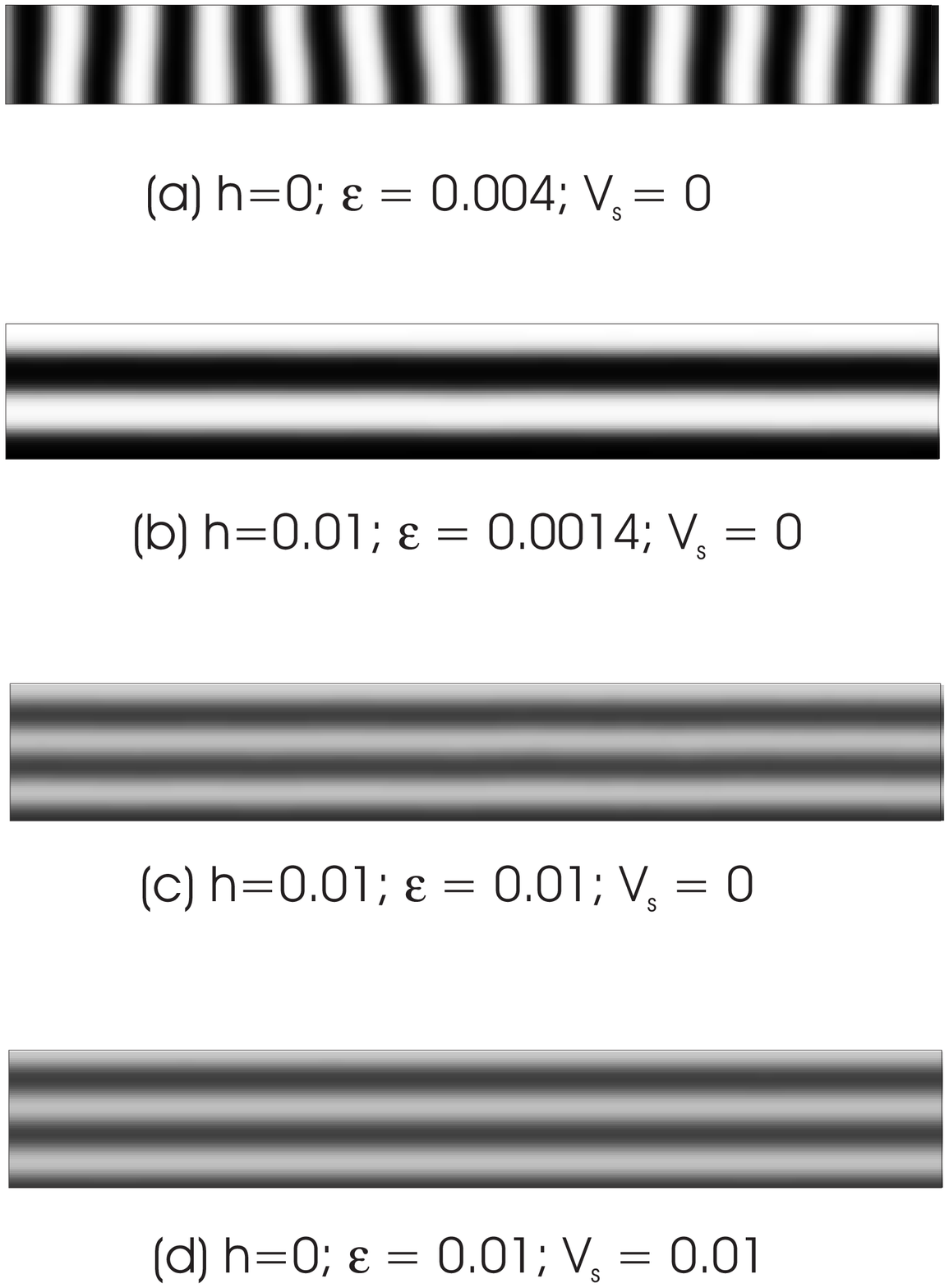}}
\end{picture}
\caption{Equilibrium patterns for confined films with $L_z$ = 21 and $L_x$
= 256 (only the first 150 columns are shown). 
(a) Neutral walls, and
matched densities. The lamellae are normal to the walls with the
bulk periodicity. 
(b) Neutral walls and density mismatch, $\epsilon$ in the SSR.
1.5 lamellae are accommodated parallel to the walls. 
(c) Neutral walls and  density mismatch, $\epsilon$ in the WSR. 2.5 lamellae 
are formed parallel to the walls. The lamellar width is 8.401, smaller then the bulk
value of 9.422.
(d)Interacting walls and matched densities, $\epsilon$ in the WSR. The lamellae are also parallel to
the walls but are more segregated than in (c).} 
\label{fig-B}
\end{figure}

Just to make sure that the bulk lamellar width as measured above was not
affected by 
the interface bending of the disordered pattern, we also measured $W_b$
in
$32\times 128$ systems with hard neutral walls and zero field, or matched densities.
 As expected
the equilibrium configuration correspond to lamellae normal to the hard
walls \cite{pibal1,kellogg1} as in Fig. \ref{fig-B}(a). $W_b$ 
was then measured using
the one-dimensional structure function for each line and finally
averaging
along the $z$ direction. The values of $W_b$ found in both
determinations
agree, so we conclude that the
excessive interface curving of the disordered pattern does not affect
the
lamellar width. Since the disordered patterns are easier to obtain, we
will
consider the lamellar width obtained from them as our bulk
 equilibrium value $W_b$. 

The results below correspond to simulations with $V_s=0$ (neutral walls),
$h\neq 0$ (mismatched densities) and $h= 0$ (matched densities), and
$V_s\neq 0$ (interacting walls) and $h= 0$. As will be seen, different
patterns  regarding the lamellae orientation appear: lamellae normal 
or parallel to the hard walls and
a mixture of both.
\vspace*{3cm}
\subsection{Neutral walls}
\begin{figure}
\setlength{\unitlength}{1mm}
\begin{picture}(90,100)(0,0)
\put(50,5){\epsfxsize=8cm\epsfbox{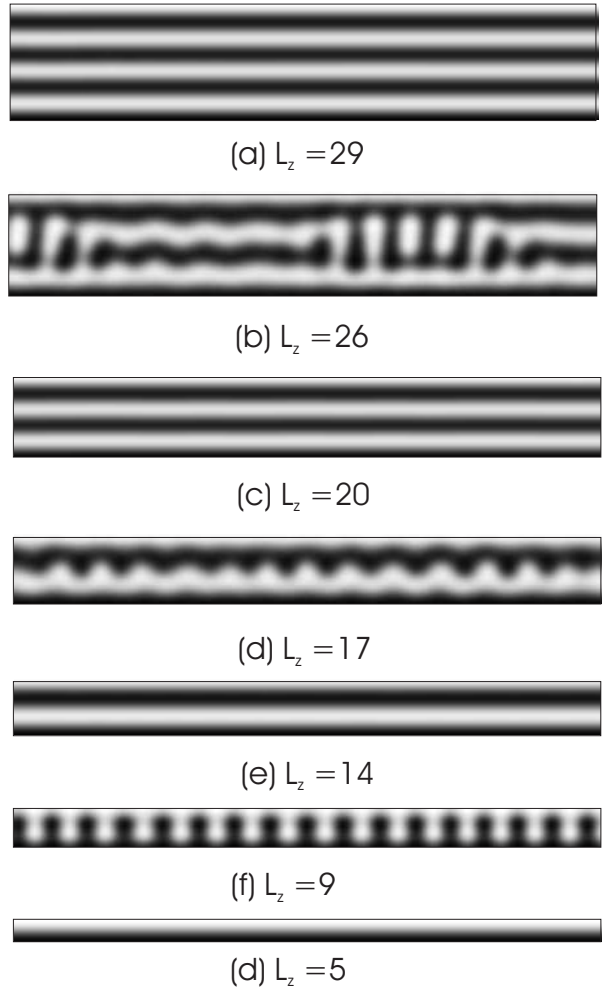}}
\end{picture}
\caption{Equilibrium patterns for films with $h=0.01$, $\epsilon=0.01$(WSR)
and variable width. $L_x=256$ but only the first 150 columns are shown.
As the width $L_z$ is decreased, the film goes
discontinuously from  $m = 3$ to $m=0$ lamellar patterns. (b), (d) and (f)
show transition patterns with mixed parallel and perpendicular lamellae.}
\label{fig-L} 
\end{figure}

 We focus now on films with mismatched densities  ($h\neq 0$), confined by
neutral walls ($V_s=0$). 
In this situation we observe patterns of lamellae parallel to the
hard walls, or a mixture of wetting layers on the hard walls and
lamellae normal to the walls in the center part of the film. First we analyze
the case of lamellae parallel to the walls only. 
Due to the density difference of chains $A$ and $B$, the denser 
part (say $A$) will be at the bottom, and the less dense  part (say
$B$), at the top. For a blend of two homopolymers $A$ and $B$, the film
would have the lower half filled with $A$ and the upper one with $B$. The
covalent bond between $A$ and $B$ parts hinders this complete separation
and forces the alternation of $A$-rich and $B$-rich microregions that will
then have
thicker interfaces due to the interpenetration of domains \cite{bamorg1}.
In the extreme case, the existence of a density mismatch may completely
destroy the segregation of $A$ and $B$.
The number of alternating lamellae  will depend on both chain size,
$\epsilon$, 
(Fig. \ref{fig-B}) and the separation between walls, $L_z$, (see Figure 
\ref{fig-L}). Also, the equilibrium
patterns will always have $m+1/2$ lamellae, where $m$=0,1,2\dots. 
\vspace*{4cm}

\begin{figure}
\setlength{\unitlength}{1mm}
\begin{picture}(90,70)(0,0)
\put(40,10){\epsfxsize=8cm\epsfbox{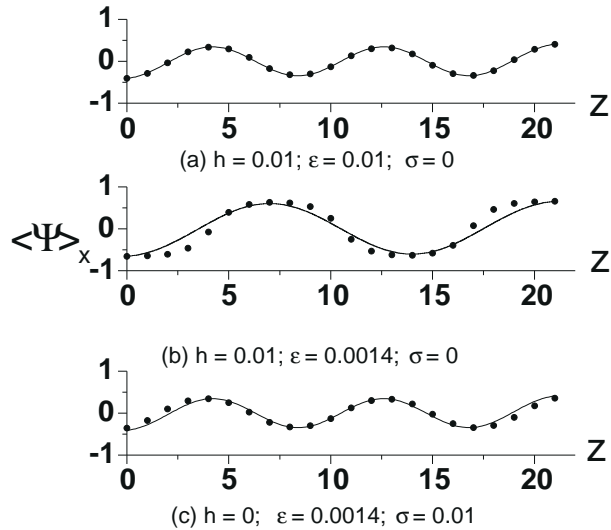}}
\end{picture}
\caption{Average concentration profiles for a film with $L_z = 21$ and $L_x = 256$.
The continuous line correspond to a fitting using Eq. (\ref{eq-profile})
(a)Neutral walls, mismatched densities. 
The profile is well fitted by a sine function plus exponential enhancement
at the hard walls. 
(b) Neutral walls, mismatched densities and molecules larger than in (a).
From the fitting it is clear
that more than one Fourier component must be considered, indicating that
the film is already in the SSR; 
(c) Interacting walls ($V_s(1) = -\sigma = -V_s(L_z)$) and matched densities.
Although similar to the profile (a), we notice that Eq. (\ref{eq-profile}) is not adequate to
describe this concentration profile.}
\label{fig-profile} 
\end{figure}

As we vary $L_z$ for a fixed $\epsilon$ we clearly see the effects of
frustration. Figure \ref{fig-L} 
shows the transitions from $m$ = 0, to $m$ = 1, $m=2$ and $m=3$
patterns  as $L_z$ is changed from 5 to 29 for $\epsilon$ = 0.01 and $h$ =
0.01. The transition patterns are frustrated and present lamellae normal
to the walls in the central region. Since full lamellar patterns are essentially one dimensional, we define the average concentration profile, 
$\langle \psi(n_z)\rangle_x$, as the average over the $x$ direction 
of the vertical variation of $\psi$. Figure \ref{fig-profile} shows the behavior of
$\langle \psi(n_z)\rangle_x$, for three different situations, for now we are interested 
in cases (a) and (b) only. 
Figure \ref{fig-profile}(a) corresponds to the profile for $\epsilon = 0.01$, $h = 0.01$, $V_s=0$ and $L_z=21$.
If we try to fit a 
sine function to that profile, we see that the fitting will miss only the wetting layers.
From this we conclude that the system is in the WSR so that the inner layers 
can be described by just one Fourier component \cite{leib1}. The wetting layers 
have an enhanced concentration due to gravitational field and the presence of the wall:
the bottom and top $AB$ layers are considerably stretched by the effect of buoyancy
and do not experiment the penetration of other layers, resulting in an excess of $A$ 
at the bottom and of $B$ at the top.

\begin{figure}
\setlength{\unitlength}{1mm}
\begin{picture}(90,70)(0,0)
\put(50,10){\epsfxsize=8cm\epsfbox{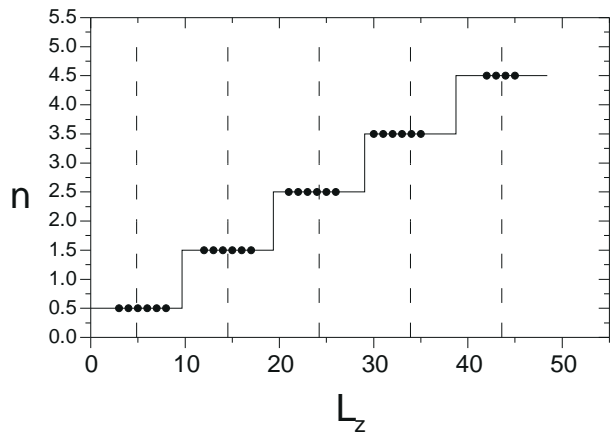}}
\end{picture}
\caption{Total number of lamellae, $n$, as a function of the film thickness $L_z$
for a film with mismatched densities, confined by neutral walls.
The solid points on the steps represent
lamellar patterns with $n$ lamellae parallel to the hard walls and between 
consecutive steps the film is in a mixed configuration with horizontal 
and vertical lamellae. 
The vertical dashed lines correspond to the 
film thickness adequate to accommodate the corresponding number of lamellae
but with the bulk width $W_b$. We see that as the film width increases,
discontinuous transitions between $n$ stretched lamellae and $n+1$ compressed
lamellae occur.}
\label{fig-m}
\end{figure}

A correction for this effect led us to the tentative function
\begin{equation}
\psi(x) = (-1)^{m+1} \eta \sin q x +2Ce^{-\beta L}\sinh 2\beta x\;\;\;\;\mbox{ for $-L/2\leq
x \leq L/2$},
\label{eq-profile}
\end{equation}
which fits very well the profile in Figure \ref{fig-profile}(a). 
For films in the WSR we found that the fitted value for $q$ is indistinguishable from $2\pi(m+1/2)/L_z$, so we define the average
lamellar width $W$, directly from the fitting, as $2\pi/q$. As will be seen below,
the interaction with the gravitational field causes a distortion within
the lamellae regarding the width of the $A$ and $B$-rich layers, but, if considered
as a unit, all the lamellae have a width very close to the average value.
 The transitions
between consecutive values of $m$ as we vary $L_z$ is shown in Figure \ref{fig-m}. 
From this 
figure we see that discontinuous transitions occur from a pattern in which the lamellae
are stretched, compared to its bulk state, to a compressed state with one more 
lamella, as $L_z$ is increased.
The regions between steps of fixed $m$ correspond to transition patterns in which
lamellae normal to the walls form in the center portion of the film.
%

\begin{figure}
\setlength{\unitlength}{1mm}
\begin{picture}(90,70)(0,0)
\put(50,5){\epsfxsize=8cm\epsfbox{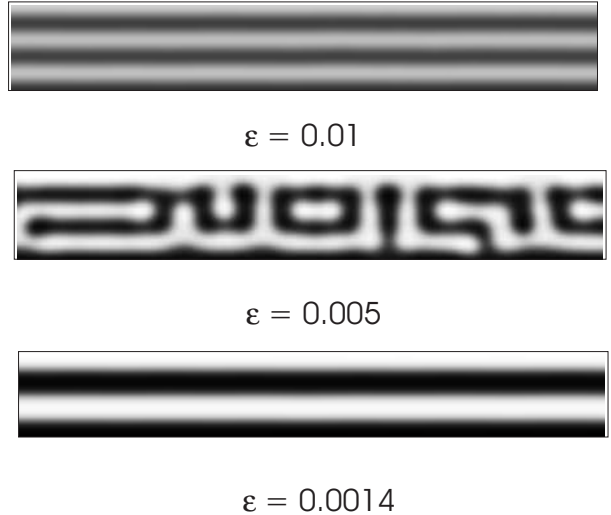}}
\end{picture}
\caption{Equilibrium patterns for confined films with $L_z$ = 21 and $L_x$=
256 (only the first 150 columns are shown) three different values of $\epsilon$.
As in the case of variable film width, the number of lamellae varies
discontinuously and transition patterns with mixed orientation lamellae
appear. For $\epsilon=0.01$ 2.5 weakly segregated lamellae are formed, 
the concentration profile can be well fitted by Eq. \ref{eq-profile}.
$\epsilon = 0.005$ produces a transition pattern and $\epsilon = 0.0014$, a
more segregated pattern with 1.5.}
\label{fig-trans} 
\end{figure}

We can, alternatively, fix $L_z$ and vary $\epsilon$, which corresponds to
fixing the film width and varying the bulk lamellae width.  
For $0.006<\epsilon<0.018$ we obtain patterns with $m=2$, in the range
$0.004  <\epsilon<0.006$ again we observe a transition pattern, and 
decreasing $\epsilon$ further  we find that a $m$ = 1
pattern appears. 
Figure \ref{fig-trans} shows patterns with $m$=1, $m$=2 and in the transition region.
The analysis of the transitions in this case is more complicated since
for $\epsilon<0.004$ the system is no longer in the WSR as can be seen
from the fitting of Eq. (\ref{eq-profile}) in Fig. \ref{fig-profile}(b). It is
clear that other Fourier components need to be included in this case.

\begin{figure}
\setlength{\unitlength}{1mm}
\begin{picture}(90,70)(0,0)
\put(50,15){\epsfxsize=10cm\epsfbox{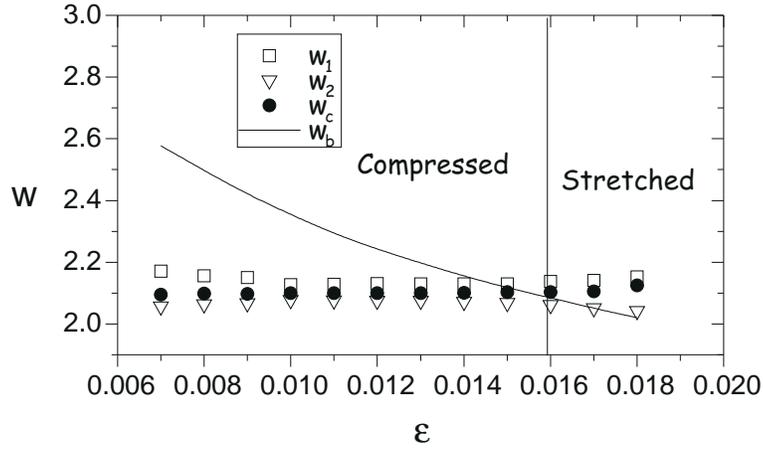}}
\end{picture}
\caption{Variation of $A$-rich and $B$-rich layers as a function of $\epsilon$ 
for a 
films with 2.5 lamellae. $w_1$, $w_2$ are the widths of the lowest $A$-rich and 
$B$-rich layers, $w_c$ is 1/4 of the central lamella and $w_b$ is 1/4 of the 
bulk lamellar width. The vertical line indicates the separation between 
regions where the film is in compressed and stretched states.}
\label{fig-we} 
\end{figure}

The accommodation of the lamellae distorts their widths non uniformly as may  be 
easily checked from the plot of the width of each individual $A$-rich and 
$B$-rich layer.
Figure \ref{fig-we} shows the behavior, as a function of $\epsilon$, of 
the widths of the first $A$-rich layer that wets the bottom wall ($w_1$), the first
$B$-rich layer connected to the bottom wetting layer ($w_2$) and  one quart
of the central lamella ($w_c = W/4$). For the sake of comparison, the variation of the 
bulk width, $w_b = W_b/4$, of $A$-rich (or $B$-rich) 
layers is also plotted. Although the variation is small compared to
the bulk behavior, we see that $w_2<w_c<w_1$ consistently.
This happens 
because there is a reduction in the number of $A$ and $B$ contacts
in the first layer (for the lack of neighboring molecules from below)
and an increase in the next one because the gravitational field 
shifts the $A$ parts downwards.
As we separate the regions where the lamellae are
compressed and stretched as compared to the bulk, we notice two
different situations.
In the compressed region, $w_1$ increases as $N$ increases ($\epsilon$ decreases),  
due to a compression of the internal layers caused by the 
greater stretching of the surface layers which produces an increase in 
the internal pressure (for $L_z$ fixed).
We expected the inverse effect to happen when $N$ was reduced in the stretched region: 
the internal layers would shrink producing a tension that would stretch the surface 
(larger effect) and second 
layers (smaller effect). But, in fact, we observe a drastic 
reduction of the second layer acting as  a tension center for
surface and central layers (Fig. \ref{fig-we}).
%

It is clear that the above effects are meaningful only for thin films. The transition
from this to the bulk behavior may be observed as we analyze $w_1$, $w_2$ and $w_c$ as 
a function of the film thickness $L_z$. If the bulk behavior prevails, $w_2\approx w_1\approx w_2\approx w_b\approx L_z/m$. As we increase $L_z$ and observe films 
with increasing number of lamellae, we find $w_1\rightarrow w_2\rightarrow w_c\rightarrow w_b$. On the other hand, the slope $\alpha$ of each group of $w$ values is proportional to $m^{-0.8}$ instead of $m^{-1}$, which reflects the different behavior of each layer of DBCP under stretching or compressions.
%
\begin{figure}
\setlength{\unitlength}{1mm}
\begin{picture}(90,60)(0,0)
\put(50,10){\epsfxsize=8cm\epsfbox{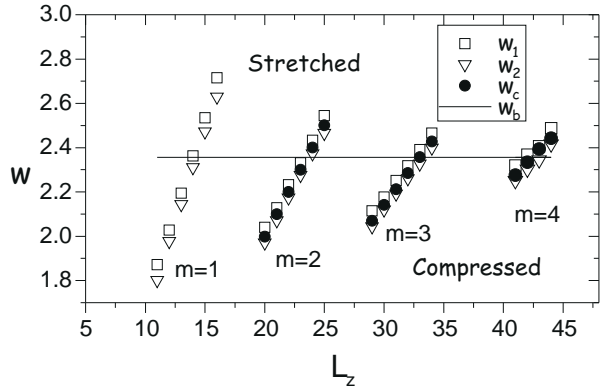}}
\end{picture}
\caption{Variation of $A$-rich and $B$-rich layers as a function of $L_z$ for a 
films with $\epsilon = 0.01$. We use here the same notation of 
Fig. \ref{fig-we}. Each group of points corresponds to lamellar patterns with 
$(m+1/2)$ lamellae. As $L_z$, and correspondingly $m$, increases, the film
behaves more like a bulk sample in the sense that the distortion of lamellae is less
significant. The slope $\alpha$ of each group is proportional to $m^{-0.8}$, for larger values
of $L_z$ a crossover to the bulk behavior $\alpha\propto m^{-1}$ is expected}
\label{fig-wL} 
\end{figure}
\subsection{Interacting walls}
The effect of surface fields in the formation of lamellar patterns has been
extensively studied \cite{turner1,kiku,brown-chakra2,pibal1,menelle,lamb,kellogg1}.
Our goal here is to compare the effect of the surface and bulk fields, so we 
consider only a film of DBCP molecules with matched densities confined by interacting walls 
in such a way that the bottom wall
attracts the denser component and the top wall prefers the less dense component,
As will seen below, in many ways this choice of walls produces a pattern 
is similar to the one obtained with neutral walls and a density mismatch, but
the two situations are, in fact, different.

The above interaction with walls may be simulated by 
choosing the surface interaction as:
$V_s = \sigma$ for $z=1$ and $V_s = -\sigma$ for $z = L_z$. The equilibrium
pattern obtained for $\epsilon = 0.01 $, $L_z = 21$, $h = 0$ and $\sigma = 0.01$ 
is very similar
to the one with the same values of $\epsilon$ and $L_z$, $\sigma = 0$ and 
$h = 0.01$, as both
present 2.5 lamellae parallel to the walls 
(see Fig. \ref{fig-B} (c) and (d), respectively). The first distinction 
appears in the segregation of domains: it is clear that the pattern in 
Fig. \ref{fig-B}(c) is less segregated due to effect of interpenetration of
domains driven by the gravitational field.
As we try to fit the concentration profile with Eq. ({\ref{eq-profile}) we
notice that the patterns with density mismatch and surface interaction are
also a little different in the surface region, so we conclude that 
Eq. (\ref{eq-profile}) is a good fit for the concentration profile of lamellar
patterns with density mismatch and neutral walls only.
A substantial difference appears for larger 
values of $h$ and $\sigma$. Figure \ref{fig-largefield} shows patterns with the 
same value of $\epsilon = 0.01$ and $L_z=21$ but one with surface field only 
and the other with density mismatch only. In this case, the lamellar 
structure still exists for $\sigma = 0.04$ but here, for $h = \sigma$ the
lamellar structure is completely destroyed in the center of the film.
%
\begin{figure}
\setlength{\unitlength}{1mm}
\begin{picture}(90,50)(0,0)
\put(50,5){\epsfxsize=8cm\epsfbox{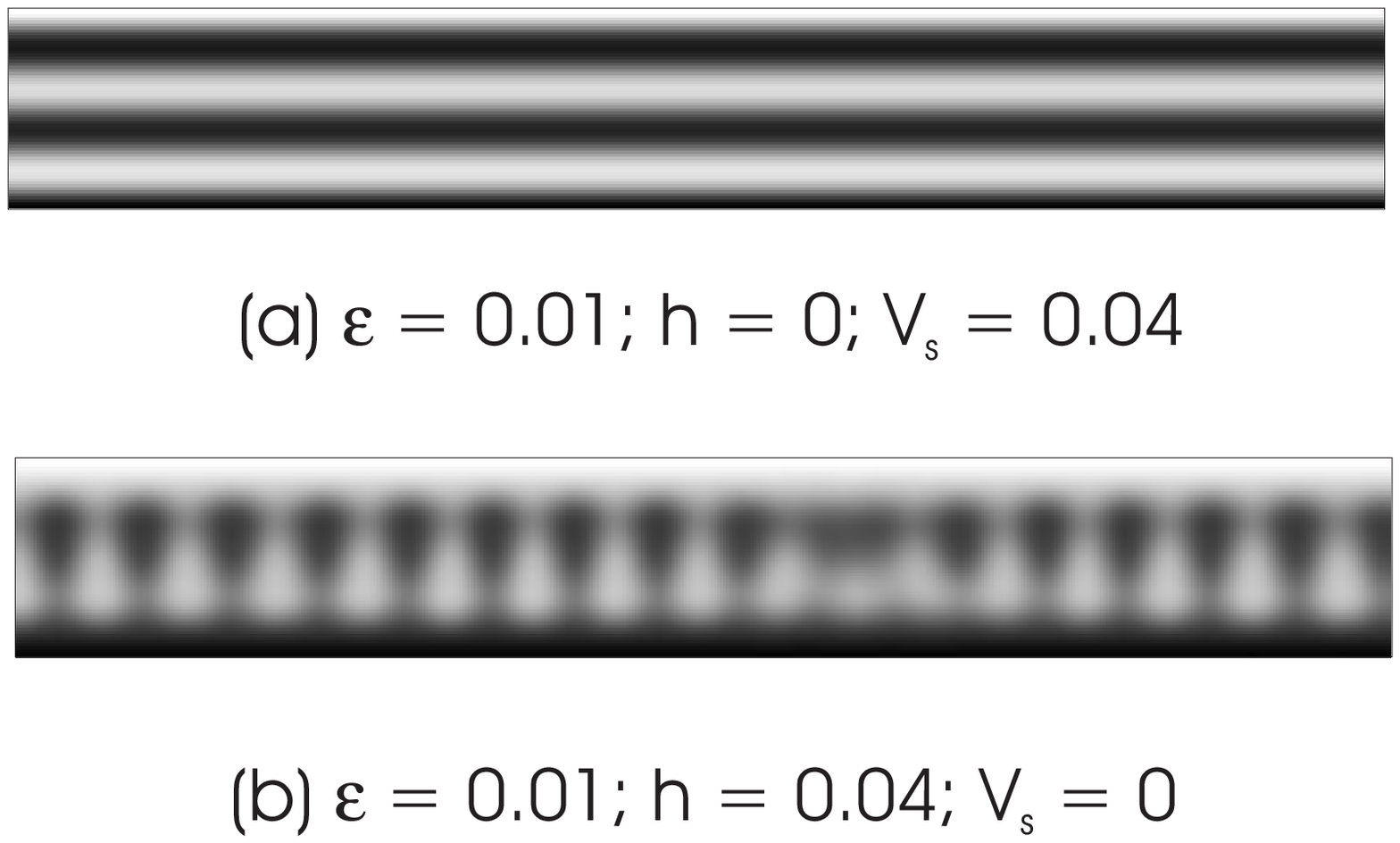}}
\end{picture}
\caption{A comparison between bulk and surface interactions.
(a) Interacting walls and matched densities. Although the surface interaction
is stronger than the one considered so far, the only noticeable difference 
is the segregation of the wetting lamellae.
(b) Neutral walls and mismatched densities. Increasing the value of the bulk
field $h$ by the same amount as the surface field the observed pattern changes dramatically:
instead of a lamellae pattern we observe a frustrated mixed orientation pattern.}
\label{fig-largefield}
\end{figure}

%
\section{Conclusions}
In this paper, we study the effects of surface and 
bulk (gravitational) fields coupled with hard wall 
restrictions on the lamellar pattern formation of 
diblock copolymer systems. We find that the two are
the predominant factors to determine the final
equilibrium pattern: lamellae tend to form 
normal to the field and their number is determined by
the ability of the system to resolve the frustration
caused by the confinement. Unresolved patterns present
a mixture of wetting lamellae normal to the field
and lamellae parallel to the field in the central part
of the film. The gravitational field also distorts
the periodicity of the lamellar pattern. The bottom $A$
layer is larger than it would be if placed in the central part
of the film. On the other hand, the next $B$ layer is
narrower, in such a way that the first lamella,
defined as the sequence $ABBA$, has a width very
close to the central lamellae.

We obtained a good fit for the 
average concentration profile in the WSR 
by using a trial function which consists of the superposition of
a sinusoidal function, characteristic of the WSR, 
and  exponential functions for the enhanced concentration
of the wetting layers.

\acknowledgements
This work was partially supported by CNPq (Brazil) and Faperj (Rio de Janeiro).

\end{document}